\documentclass[12pt]{elsarticle}
\graphicspath{{Figures/}}
\usepackage{amsmath,amssymb,amsthm,amsfonts}
\usepackage{caption}
\usepackage[title]{appendix}
\usepackage{xcolor}
\usepackage[normalem]{ulem}
\setcitestyle{authoryear,open={(},close={)}}

\usepackage{graphicx,epsfig}
\usepackage{psfrag}
\usepackage{xcolor}
\usepackage[hmargin=1in,vmargin=1in]{geometry}
\usepackage{stabular,morefloats}
\usepackage{textcomp}
\usepackage{titlesec}
\usepackage{colortbl}
\usepackage{xcolor}
\usepackage{multirow}
\usepackage{mathrsfs}
\usepackage{tabularx}
\usepackage{comment}
\usepackage[colorlinks]{hyperref}
\usepackage{multirow,tabularx,afterpage,fancyhdr}
\usepackage{placeins}
\titleformat{\subsubsection}
  {\normalfont\selectfont}{\thesubsubsection}{1em}{}
\numberwithin{equation}{section}
\usepackage{float}
\usepackage{xcolor}
\usepackage{soul}
\usepackage{cleveref}
\definecolor{lightblue}{rgb}{0,0,1}
\sethlcolor{lightblue}

\usepackage[subrefformat=parens,labelformat=parens]{subfig}

\colorlet{Changes@Color}{black}
\makeatletter
\renewcommand{\@biblabel}[1]{[#1]\hfill}
\makeatother

\usepackage{etoolbox}
\makeatletter
\patchcmd{\ps@pprintTitle}
  {Elsevier}
  {The Journal of the Mechanics and Physics of Solids}
  {}{}
\makeatother

\begin{document}

\sloppy

\title{Modeling damage and fracture in additively manufactured polymeric triply periodic minimal surface lattices}

\author[main]{Abhishek Gupta\fnref{equal}}
\author[main]{Aditya Konale\fnref{equal}}
\author[main4]{Ke Ma}
\author[main]{Keven Alkhoury}
\author[main]{Pradeep Guduru}
\author[main3,main4]{Yuri Bazilevs}
\author[main,main2]{Vikas Srivastava\corref{cor1}}
\ead{vikas\_srivastava@brown.edu}
\cortext[cor1]{Corresponding author}
\fntext[equal]{These authors contributed equally to this work.}
\address[main]{School of Engineering, Brown University, Providence, RI 02912, USA}
\address[main2]{Institute for Biology, Engineering and Medicine, Brown University, Providence, RI 02912, USA}
\address[main3]{Department of Mechanical Engineering, Vanderbilt University, Nashville, TN 37235, USA}
\address[main4]{Department of Civil and Environmental Engineering, Vanderbilt University, Nashville, TN 37212, USA}
%\address[equal]{These authors contributed equally to this work.}

\begin{abstract}

Architected triply periodic minimal surface (TPMS) lattices offer superior specific energy absorption, toughness, fatigue strength, and tunability. While recent advancements have established rate-dependent viscoplastic constitutive models to capture the complex nonlinear deformation response of additively manufactured polymeric TPMS structures, predicting fracture and the resulting structural failure remains a significant challenge. We address this by performing systematic experiments on unit cells and lattices of various sizes under tension, compression, and non-monotonic loading. The experiments inform the development of a new constitutive model that captures the damage and fracture behavior of polymeric TPMS lattices. We first implement a high-fidelity viscoplastic deformation constitutive model from \citet{ma2026polymer} into finite element software Abaqus/Explicit via a user material subroutine. We then propose a damage initiation criterion for amorphous polymers based on stored elastic energy and equivalent plastic strain. The damage model is implemented in Abaqus using gradient-damage framework following \citet{konale2025modeling}. The damage model and numerical simulation capability are quantitatively and qualitatively validated using experimental results for a unit cell under non-monotonic loading and lattices under tension. The proposed damage model and simulation capability enable in silico design of architected polymer structures.

\end{abstract}

\vspace{0.25in}


\maketitle

\section{Introduction}

Triply Periodic Minimal Surfaces (TPMS) are mathematically defined geometries characterized by zero mean curvature and the ability to periodically tessellate in all three spatial dimensions \citep{schoen1970infinite,qiu2024experimental,feng2022triply,yu2019investigation}. Additively manufactured TPMS lattices and self-assembled spinodal metamaterials, have gained significant attention due to their smooth topologies, which minimize stress concentrations \citep{li2025additively,kumar2020inverse,al2019multifunctional}. This unique morphology enables exceptional stress uniformity \citep{preve2023comprehensive,al2018topology}, fatigue resistance \citep{speirs2017fatigue,cai2025overcoming},  toughness \citep{zhao2025parametric}, recoverability \citep{wan2025energy,portela2020extreme}, energy absorption \citep{al2018topology}, and relatively isotropic mechanical properties \citep{chatzigeorgiou2022numerical}, outperforming conventional strut- and plate-based architected lattices \citep{chatzigeorgiou2022numerical,al2018topology}.
Among TPMS topologies, sheet-based lattices have been shown to exhibit stretch-dominated behavior \citep{abou2020mechanical}, resulting in linear density-dependent scaling of mechanical properties such as elastic modulus and compressive strength, similar to that observed in Octet-truss lattices \citep{al2018topology,kadkhodapour2017relationships}. The more uniform stress distribution within TPMS unit cells delays the onset of failure compared to the stress concentrations that develop in strut-based lattices \citep{bauer2016approaching}, resulting in toughness and compressibility that rival those of bending-dominated architectures \citep{al2018topology}. These exceptional mechanical properties, combined with the tunable open-cell morphology and high surface-area-to-volume ratio of TPMS lattices, have made them attractive for a wide range of applications, including structural engineering \citep{lin2022mechanical}, orthopedic implants \citep{yan2015ti}, tissue regeneration scaffolds \citep{afshar2016additive}, heat exchangers \citep{iyer2022heat}, phononics \citep{hur2017exploring}, and impact-protection systems \citep{zhao2025parametric}. 
However, the widespread adoption of TPMS lattices in these engineering applications requires addressing several challenges, including the establishment of scalable fabrication methods, robust structure-property relationships,  biocompatibility, environmental stability, and the prediction and mitigation of failure under complex loading conditions. We combine experiments and finite element simulations to elucidate and predict the deformation and failure behavior of Schwarz Primitive TPMS lattices across a range of lattice sizes and loading conditions, addressing a critical gap in the existing literature.

TPMS lattices have been fabricated using a wide range of additive manufacturing techniques \citep{al2018topology,afshar2016additive,ozturk2026numerical,ma2026polymer,zhao2025parametric}, and materials, including polymers \citep{ma2026polymer}, metals \citep{al2018topology,qiu2023mechanical}, and ceramics \citep{lu2024enhancing}, across various length scales \citep{maldovan2007sub,harley20253d}. Powder-bed fusion processes, such as selective laser melting (SLM) of metals and selective laser sintering (SLS) of polymers, produce lattices with relatively isotropic mechanical properties that generally exhibit high toughness, as the constituent materials are inherently ductile and damage tolerant \citep{qiu2024experimental,zhao2025parametric}. Consequently, their fracture behavior is often less sensitive to geometric parameters, such as unit-cell topology and relative density, as well as testing conditions, including strain rate \citep{zhao2025parametric,al2018topology}. In contrast, lattices fabricated using fused deposition modeling (FDM) typically exhibit anisotropic mechanical properties and are more susceptible to voids and manufacturing defects \citep{yuan2026quasi,ozturk2026numerical}. In this work, we employ stereolithography (SLA), an additive manufacturing process that produces structures with relatively isotropic mechanical properties but limited toughness due to the thermosetting nature of the resin, resulting in comparatively brittle polymer lattices \citep{ma2026polymer,yu2019investigation,melchels2010mathematically}. We demonstrate that a strain-rate-dependent brittle-to-ductile transition in fracture occurs which is also a function of printed layer thickness and relative density of the structure. This transition gives rise to a rich rate-dependent constitutive and failure response, providing a unique opportunity to investigate and model the mechanisms governing damage initiation, damage evolution, and ultimately fracture. TPMS lattices have been extensively characterized under quasistatic compression, revealing the influence of topology \citep{yuan2026quasi,al2018topology}, cell size \citep{maskery2017compressive}, density gradients \citep{yu2019investigation,afshar2016additive,qiu2023mechanical}, and composite reinforcement on energy absorption \citep{ozturk2026numerical}, elastic modulus, and compressive strength. Their dynamic behavior has also been investigated using direct-impact Hopkinson bar \citep{novak2023high,yuan2026quasi}, Kolsky bar \citep{santiago2023modelling}, dynamic mechanical analysis \citep{cai2025overcoming}, and drop-weight tests \citep{zhao2025parametric}, which have shown inertia-induced strengthening and the resulting enhancement in energy absorption. 
Although constitutive models have been developed for TPMS lattices, most are limited to describing the deformation-only behavior of relatively tough lattices fabricated using SLM and SLS under compressive loading \citep{abou2020mechanical,qiu2024experimental,abueidda2019mechanical}. Furthermore, these models are often restricted to a single lattice size or loading condition. 
While advanced computational frameworks are being developed to track progressive damage in  elastomers \citep{MAO2018,MOUSAVI2025,konale2026physics}, glassy polymers \citep{narayan2021fracture,SOHRABIZADEH2025}, polymer-matrix composites \citep{HASSAN2008,COZAR2022,Vaishakh2024,Weican2024},  applying such computational approaches to predict  structural failure in additively manufactured polymeric TPMS remains an open challenge. 
Currently, the modeling of damage initiation, crack nucleation, and failure has largely been confined to either limited computational studies \citep{preve2023comprehensive} or experiments conducted under compressive loading at fixed strain rates \citep{qiu2026mechanical}, leaving a significant gap in understanding the failure behavior of rate-dependent polymer-based TPMS lattices under complex loading conditions and across varying unit-cell tessellations.

We use the finite element framework in Abaqus/Explicit and incorporate an appropriate damage initiation criterion together with a gradient-damage formulation into a rate-dependent viscoplastic constitutive model for polymeric material to capture and accurately predict the deformation and failure behavior observed experimentally in additively manufactured TPMS lattices. The following provide a brief outline of the study and summarize the key contributions of this paper.

\begin{enumerate}[i]

\item Our uniaxial compression experiments on $4\times4\times4$ lattices revealed that a smaller printing layer thickness and lower relative density provide the highest resistance to brittle fracture under high-strain-rate loading. This finding informed the selection of a $25~\mu\mathrm{m}$ layer thickness and a $5~\%$ relative density for the present study.

\item We first implemented the deformation model from \citet{ma2026polymer} in the FE software Abaqus/Explicit through a user material subroutine. We then developed a damage initiation criterion for amorphous polymers based on stored elastic energy and equivalent plastic strain by extending the fracture framework of \citet{konale2025modeling}. This damage criterion is then integrated into a damage and fracture modeling framework that accounts for damage initiation and damage evolution through a gradient-damage formulation which was implemented within Abaqus using VUMAT and VUEL.

% \item We then implement a deformation-only constitutive model \citep{ma2026polymer} in Abaqus/Explicit via user-defined subroutines to simulate dog-bone tensile experiments, single-unit-cell compression experiments, and the compressive response of $4\times4\times4$ TPMS lattices.

% \item Following validation of the constitutive response, we develop a damage initiation criterion for amorphous polymers based on stored elastic energy and equivalent plastic strain by extending the fracture framework of \cite{konale2025modeling}. This criterion is subsequently integrated into a damage and fracture modeling framework that accounts for damage initiation and evolution through a gradient-damage formulation implemented within Abaqus.

\item The predictive capability of the damage constitutive model and numerical framework is assessed through quantitative and qualitative comparisons with experiments on a TPMS unit cell subjected to uniaxial non-monotonic loading and TPMS lattices under uniaxial tension. The proposed fracture modeling approach offers mechanistic insights into the evolution of deformation and failure across multiple structural scales and facilitates the \textit{in silico} design of polymer-based architected materials and structures.
 
\end{enumerate}

%Superior performance beyond the limits of bulk materials can be achieved by exploiting topology in architected structures \citep{Architected1, Architected2, Architected3}. Among these, triply periodic minimal surface (TPMS) designs are particularly promising due to their smooth geometry, high connectivity, and inherent tunability \citep{qiu2024experimental, feng2022triply, yu2019investigation}. Utilizing polymers as a base material offers additional advantages, including low density, environmental stability, and biocompatibility \citep{PolymerTPMS1}. 

%Damage and fracture modeling of polymeric TPMS structures is also important from both scientific (for physics-based insights) and practical application (for completeness of the design process) perspectives. However, there is a significant corresponding research gap in the literature. The finite element method (FEM) with gradient-damage is a suitable numerical framework for this purpose \citep{konale2025modeling}. We make the following contributions here: 

\section{Fabrication and Experiments}

%\subsection{TPMS unit cell geometry, fabrication, and experimental setup}

\begin{figure}[ht!]
    \begin{center}
		\includegraphics[width=0.9\textwidth]{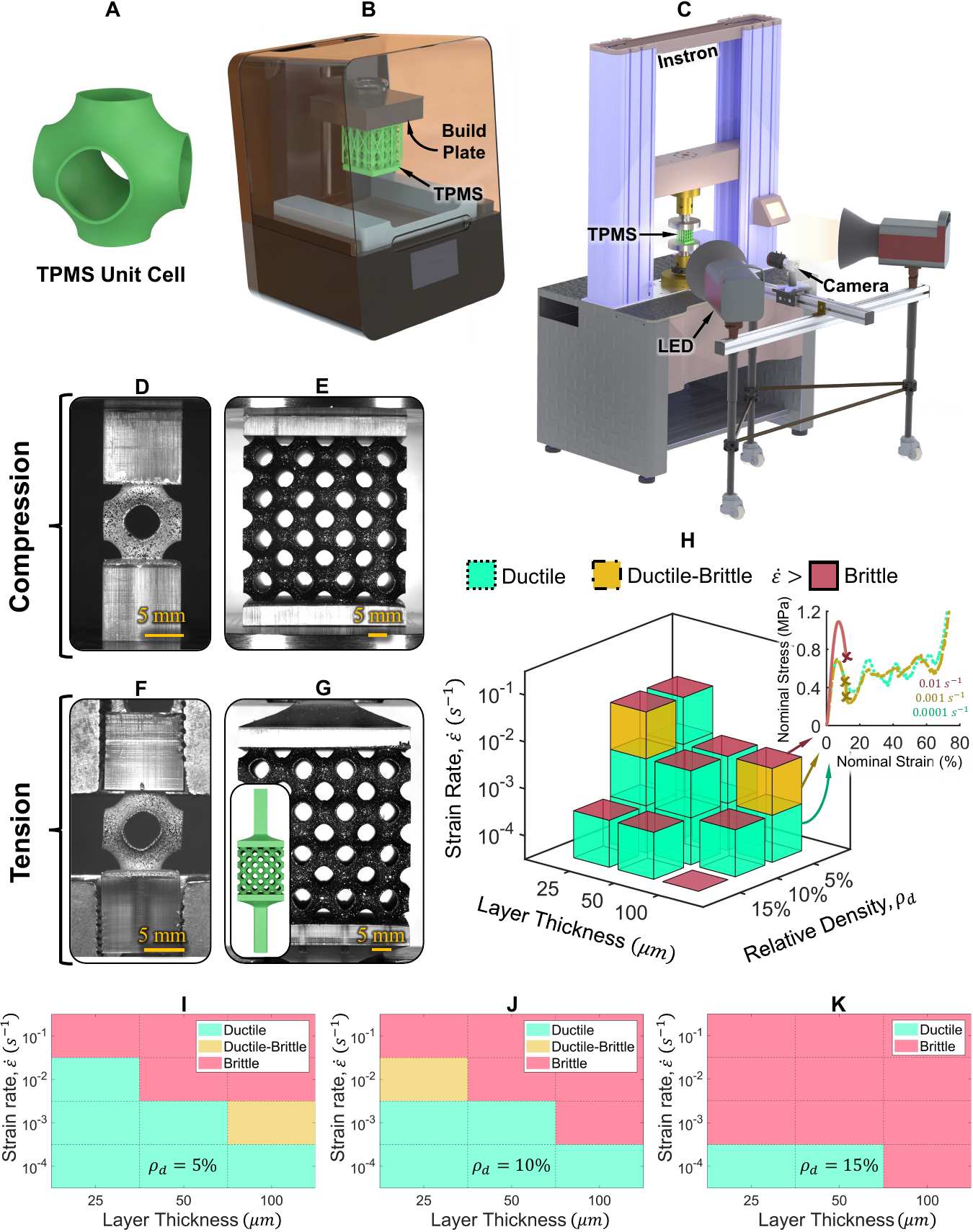}
	\end{center}
    \caption{\small {\textbf{Schwarz-P TPMS unit cell geometry and experimental setup.} \textbf{(A)} TPMS unit cell geometry. \textbf{(B)} SLA 3D printer for fabrication. \textbf{(C)} Experimental setup for compression, tension, and in-situ imaging. Pictures of single unit cell and $4\times4\times4$ lattice in \textbf{(D, E)} compression and \textbf{(F, G)} tension. \textbf{(H)} Parameter space demonstrating brittle to ductile transition. \textbf{(I, J, K)} 2D plots showing the brittle-to-ductile transition, presented separately for each relative density. }} 
    \label{TPMSgeomfig1}
\end{figure}

\Cref{TPMSgeomfig1}(A) illustrates the geometry of a Schwarz-P TPMS unit cell. Lattices of different sizes were generated by tessellating this single unit cell in three dimensions. \Cref{TPMSgeomfig1}(B) shows a \mbox{Formlabs Form 3+} stereolithography (SLA) 3D printer utilized for fabricating all samples experimentally characterized in this study. We used Formlabs Clear Resin V4, a thermoplastic polyolefin (TPO)-based translucent UV-curable resin with high post-curing stiffness, enabling structural integrity and high dimensional accuracy during printing with minimal support structures. The lattice geometrical designs were the same as in \citet{ma2026polymer} and were optimized to mitigate resin pooling and bulging effects arising from surface tension. Solid end blocks were added to two opposite sides of the samples in the CAD design along the intended deformation direction to ensure planar loading and laterally fixed boundary conditions [\Cref{TPMSgeomfig1}(D to G)]. These end blocks also helped prevent damage to the lattice structure during support removal, washing, and post-curing. Although SLA printing is expected to produce samples with isotropic properties, we printed all samples with the printing direction aligned with the intended deformation direction during mechanical testing for consistency. Once printed, the samples were washed in isopropyl alcohol (IPA) using mechanical agitation in a Form Wash for 10 minutes and were then left in the open for one hour to allow excess IPA to evaporate. After drying, we removed the supports and post-cured the samples in a Form Cure under 405 nm UV light at $60^\circ$C for 15 min for the single-unit-cell samples and $2\times2\times2$ lattices, and for 30 min for the $4\times4\times4$ lattices because of their larger volume. Once fully cured and cooled, we measured the sample dimensions using a digital vernier caliper and measured the total mass (lattice/unit cell + end blocks) using a microbalance. After mechanical testing, we scraped the compressed and fractured lattices/unit cells from the end blocks, measured the mass of the end blocks individually for each sample, and subtracted it from the previously measured total mass to obtain the mass of the lattice/unit cell. We divided the mass of each sample by its total bounding-box volume to calculate the bulk density $(\rho)$. We then normalized this value by the density of the solid polymer, $\rho_s = 1181.8 \pm 3.5~\mathrm{kg/m^3}$, to obtain the relative density $(\rho_d=\rho/\rho_s)$. The average relative density of the unit-cell samples was $5.9 \pm 0.8~\%$, which is close to the target value of $5~\%$ in the CAD design. The relative densities of the $2\times2\times2$ and $4\times4\times4$ lattices were $6.5 \pm 0.1~\%$ and $7.0 \pm 0.4~\%$, respectively, slightly higher than the target density of $5~\%$ due to their more intricate geometries.

\Cref{TPMSgeomfig1}(C) shows an illustration of the Instron 5900R 5882 with a $100~\mathrm{kN}$ load cell used for mechanical testing of samples under both tension and compression. We built a custom mobile imaging rig that can be attached to the Instron that included a camera (Basler acA5472-17um) mounted on a linear rail to adjust the field of view for imaging samples of various sizes and LED lamps for illumination. \Cref{TPMSgeomfig1}(D, F) shows 3D-printed single-unit-cell samples in compression and tension configurations. We kept the unit-cell design identical for both compression and tension experiments. The solid end blocks in the unit cells were sufficiently small to fit within the tensile grips during tension testing. For the $4\times4\times4$ lattices, the compression specimens include thin, flat end blocks at both ends [\Cref{TPMSgeomfig1}(E)]. For the tension specimens, extruded gripping sections were added [inset of \Cref{TPMSgeomfig1}(G)] and connected to the end blocks through necked regions. This design promotes planar loading and reduces stress concentrations in the end blocks, minimizing the likelihood of premature failure at the grips.

Our compression testing of $4\times4\times4$ lattices revealed a coupled process-structure-property relationship. In \Cref{TPMSgeomfig1}(H), each block represents a unique combination of strain rate, printing layer thickness, and relative density. Green blocks indicate combinations that produce ductile compressive behavior, where the lattice remains fully deformable up to the densification regime. Yellow blocks indicate combinations that produce near-ductile behavior accompanied by localized brittle failure. For each combination of layer thickness and relative density, the upper surface of the highest green or yellow block is colored red, marking the strain-rate threshold beyond which the lattice exhibits premature brittle failure. Representative stress-strain\footnote{The uniaxial dogbone specimen results are reported using true stress and true strain, while the unit-cell and lattice structure results are presented using engineering stress and engineering strain.} curves as a function of strain rate for $\rho_d=5~\%$ and a printing layer thickness of $100~\mu\mathrm{m}$ illustrate the distinction between ductile and brittle behavior. The green curve exhibits full deformability into the densification regime. The yellow curve exhibits localized brittle failure, appearing as a small sudden stress drop (marked by two crosses on the curve), after which the lattice continues to deform into the densification regime. In contrast, the red curve exhibits premature catastrophic brittle failure, marked by a cross at the point where the curve terminates. For the brittle failure case, the stress–strain curve terminates around $\epsilon\approx11.4\pm3.8\%$ immediately following the first stress peak. Catastrophic failure then causes the stress to drop abruptly to zero. \Cref{TPMSgeomfig1}(H, I, J, K) demonstrates a brittle-to-ductile transition as the printing layer thickness, relative density, and strain rate are reduced \citep{ma2026polymer,meza2014strong}. For the largest printing layer thickness of $100~\mu\mathrm{m}$ and a relative density of $15~\%$, the lattice fails prematurely even at the lowest strain rate ($0.001~s^{-1}$). In contrast, for the smallest layer thickness of $25~\mu\mathrm{m}$ and a relative density of $5~\%$, the lattice exhibits ductile compressive behavior up to a strain rate of $0.01~s^{-1}$ and fails prematurely only at the higher strain rate of $0.1~s^{-1}$.
Based on these observations, all samples used in this study for both tension and compression experiments were fabricated with a prescribed relative density of $5~\%$ and a printing layer thickness of $25~\mu\mathrm{m}$. This combination produced a strain-rate-dependent stress–strain response characterized by high deformability up to the densification strain and a stable, non-zero plateau stress across the widest accessible strain-rate range, providing a suitable experimental basis for constitutive modeling.

%Although applications of TPMS lattices typically involve compression, tensile experiments, which are seldom performed, are necessary for a comprehensive characterization and understanding of their response. Three repetitions were performed for each experimental result reported. Interested readers can refer to \citep{ma2026polymer} for further details. The \emph{Appendix} presents the procedure for STL file conversion for FEM.
%along with the experimental setup for unit cells and lattices ($4\times4\times4$, $2\times2\times2$) subjected to compression and tension \citep{ma2026polymer}. The 5\% target relative density for the unit cells and lattices from \cite{ma2026polymer} was considered. Solid platens on the two opposite faces normal to the printing direction were added for both the unit cell and lattice to facilitate planar loading and to impose fixed boundary conditions at the top and bottom. The lattice fixtures were redesigned for tension experiments (see \Cref{TPMSgeomfig1}(C)) to ensure that failure occurs within the lattice. 

%\setcounter{figure}{0}

\section{Constitutive Modeling}

\subsection{Implementation of the rate-dependent viscoplastic deformation constitutive model for amorphous polymers from \citet{ma2026polymer} in FEM}

The deformation model employed here is a three-dimensional, rate-dependent, finite-deformation viscoplastic constitutive model for amorphous polymers adopted from \citet{ma2026polymer}. The rate of deformation tensor is additively decomposed into elastic and plastic parts as
\begin{equation}
    \mathbf{D} = \mathbf{D}^{e} + \mathbf{D}^{p},
    \qquad
    \dot{\boldsymbol{\sigma}} = \boldsymbol{\mathcal{C}}^{e}:\mathbf{D}^{e},
    \qquad
    \mathbf{D}^{p} = \sqrt{\frac{3}{2}}\,\dot{\bar{\varepsilon}}^{p}\mathbf{N},
\end{equation}
where $\boldsymbol{\sigma}$ is the Cauchy stress, $\boldsymbol{\sigma}_{\mathrm{dev}}$ denotes its deviatoric part, $\boldsymbol{\mathcal{C}}^{e}$ is the isotropic elasticity tensor, $\dot{\bar{\varepsilon}}^{p}$ is the equivalent plastic strain rate, and $\mathbf{N}=\sqrt{3/2}\,\boldsymbol{\sigma}_{\mathrm{dev}}/\sigma_{\mathrm{vm}}$ is the associative flow direction with $\sigma_{\mathrm{vm}}=\sqrt{3/2}\|\boldsymbol{\sigma}_{\mathrm{dev}}\|$. The rate dependence is introduced through a modified power-law relation involving a hyperbolic sine function,
\begin{equation}
    \dot{\bar{\varepsilon}}^{p}
    =
    \dot{\varepsilon}_{0}
    \left[
        \sinh\left(\frac{\sigma_{\mathrm{vm}}}{S}\right)
    \right]^{1/m},
\end{equation}
where $\dot{\varepsilon}_{0}$ is a reference strain rate, $m$ is the rate-sensitivity exponent, and $S$ is the effective resistance to plastic flow consisting of three mechanisms for capturing the full rate-dependent response, labeled here as $S^{\mathrm{a}}$, $S^{\mathrm{b}}$, and $S^{\mathrm{c}}$, giving
\begin{equation}
    S = S^{\mathrm{a}} + S^{\mathrm{b}} + S^{\mathrm{c}}.
\end{equation}
Here, $S^{\mathrm{a}}$ represents a stress-like transient flow resistance associated with deformation-induced disordering and is coupled to an order parameter $\phi$ \citep{srivastava2010thermally}; $S^{\mathrm{b}}$ captures the pseudo-elastic small-strain response and subsequent saturation \citep{boyce1988large}; and $S^{\mathrm{c}}$ captures material strain hardening due to accumulated plastic deformation \citep{srivastava2010thermo} and the equivalent plastic strain rate. In a semi-implicit stress update, a trial stress $\boldsymbol{\sigma}_{n+1}^{\mathrm{tr}}$ is first computed elastically, followed by the solution of the scalar nonlinear equation
\begin{equation}
    \sigma_{\mathrm{vm},n+1}^{\mathrm{tr}}
    -
    3\mu\Delta\bar{\varepsilon}^{p}_{n+1}
    -
    \sinh^{-1}
    \left[
        \left(
            \frac{\Delta\bar{\varepsilon}^{p}_{n+1}}
                 {\Delta t\,\dot{\varepsilon}_{0}}
        \right)^{m}
    \right]
    S_{n+1}
    =
    0,
\end{equation}
where $\mu$ is the shear modulus and $\Delta\bar{\varepsilon}^{p}_{n+1}=\Delta t\,\dot{\bar{\varepsilon}}^{p}_{n+1}$. The stress and plastic strain are then updated as
\begin{equation}
    \boldsymbol{\sigma}_{n+1}
    =
    \boldsymbol{\sigma}_{n+1}^{\mathrm{tr}}
    -
    \sqrt{6}\mu\Delta\bar{\varepsilon}^{p}_{n+1}\mathbf{N}_{n+1}^{\mathrm{tr}},
    \qquad
    \boldsymbol{\varepsilon}^{p}_{n+1}
    =
    \boldsymbol{\varepsilon}^{p}_{n}
    +
    \sqrt{\frac{3}{2}}\Delta\bar{\varepsilon}^{p}_{n+1}\mathbf{N}_{n+1}^{\mathrm{tr}}.
\end{equation}
The internal variables are updated consistently with the same plastic increment according to
\begin{align}
    S^{\mathrm{a}}_{n+1}
    &\approx
    S^{\mathrm{a}}_{n}
    +
    H^{\mathrm{a}}_{0}
    \left\{
        b\left[
            z\left(
                \frac{\Delta\bar{\varepsilon}^{p}_{n+1}}
                     {\Delta t\,\dot{\varepsilon}^{\mathrm{a}}_{0}}
            \right)^{s}
            -
            \phi_{n}
        \right]
        -
        S^{\mathrm{a}}_{n}
    \right\}
    \Delta\bar{\varepsilon}^{p}_{n+1},
    \\
    \phi_{n+1}
    &\approx
    \phi_{n}
    +
    g
    \left[
        z\left(
            \frac{\Delta\bar{\varepsilon}^{p}_{n+1}}
                 {\Delta t\,\dot{\varepsilon}^{\mathrm{a}}_{0}}
        \right)^{s}
        -
        \phi_{n}
    \right]
    \Delta\bar{\varepsilon}^{p}_{n+1},
    \\
    S^{\mathrm{b}}_{n+1}
    &\approx
    S^{\mathrm{b}}_{n}
    +
    H^{\mathrm{b}}_{0}
    \left(
        1-\frac{S^{\mathrm{b}}_{n}}{S^{\mathrm{b}}_{\infty}}
    \right)
    \Delta\bar{\varepsilon}^{p}_{n+1},
    \\
    S^{\mathrm{c}}_{n+1}
    &\approx
    S^{\mathrm{c}}_{n}
    +
    H^{\mathrm{c}}_{0}
    \bar{\varepsilon}^{p}_{\mathrm{eff},n}
    (\Delta t)^{1-l}
    \left(\Delta\bar{\varepsilon}^{p}_{n+1}\right)^{l}.
\end{align}
Here, $H^{\mathrm{a}}_{0}$, $H^{\mathrm{b}}_{0}$, $H^{\mathrm{c}}_{0}$, $S^{\mathrm{b}}_{\infty}$, $b$, $g$, $z$, $s$, $l$, and $\dot{\varepsilon}^{\mathrm{a}}_{0}$ are material parameters controlling the evolution rates, saturation behavior, rate dependence, and hardening/softening characteristics of the internal variables. This condensed formulation is the constitutive basis of the Abaqus/VUMAT implementation described below.

\begin{figure}[ht!]
    \begin{center}
		\includegraphics[width=0.8\textwidth]{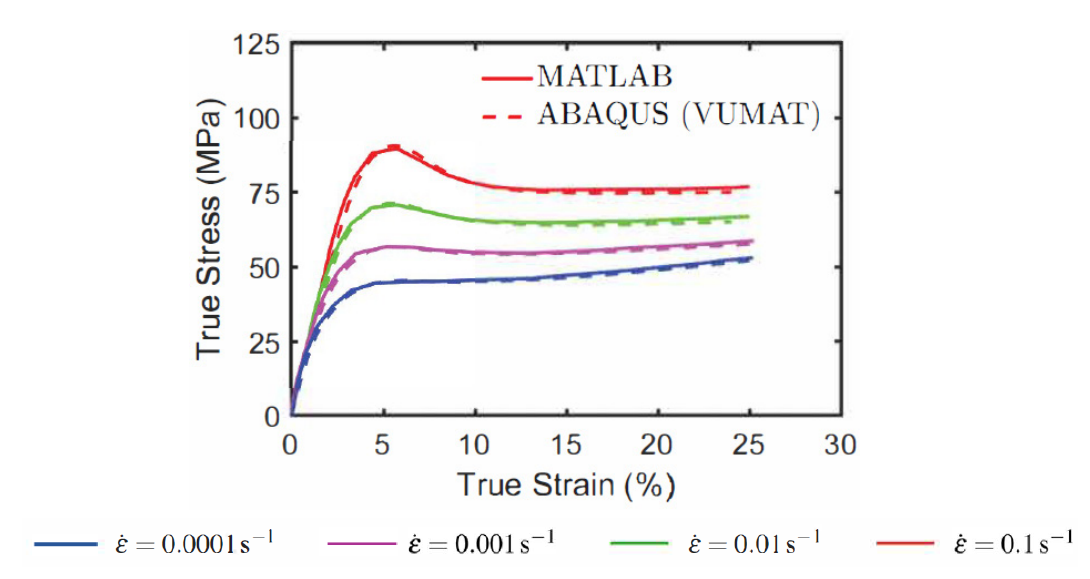}
	\end{center}
    \caption{\small {\textbf{Uniaxial tension of dogbone specimens: Comparison of single element FEM predictions with baseline Matlab results.} Experimental results are shown in \Cref{SWecrfig5}. %Good agreement of single element (C3D8R) FEM true stress-strain response predictions with baseline Matlab results from \cite{ma2026polymer} over four decades of strain rates. The following MSFs were used - 0.1 s$^{-1}$: MSF = $10^{6}$, 0.01 s$^{-1}$: MSF = $10^{8}$, 0.001 s$^{-1}$: MSF = $10^{10}$, 0.0001 s$^{-1}$: MSF = $10^{12}$). The predictions are stable and accurate even for very large MSFs, highlighting the model and the implementation's robustness.
    }} 
    \label{singleelfig2}
\end{figure}

%As the first step towards developing a FEM-based damage and fracture modeling tool for polymeric TPMS lattices, we implemented the rate-dependent viscoplastic deformation-only constitutive model for amorphous polymers from \citep{ma2026polymer} in Abaqus. 
The \emph{Appendix} presents the procedure for STL file conversion for unit cells and lattices to an independently meshable solid body in Abaqus. The model was implemented using the explicit time integration scheme via a material subroutine (VUMAT). The FEM implementation was comprehensively validated by comparing single element (uniaxial tension of dogbone specimens) and multi-element (unit cell and lattice under uniaxial compression) deformation response predictions with baseline Matlab, isogeometric analysis (IGA) \citep{ma2024isogeometric,ALAYDIN2022}, and experimental results from \citet{ma2026polymer}. Uniform wall thickness was considered for all the unit cell and lattice FEM simulations in this work. The respective bounding dimensions were used for unit cell and lattice nominal stress and strain calculations. \Cref{singleelfig2} shows excellent agreement of the single element (using C3D8R, three-dimensional 8-node linear brick element with reduced integration) FEM true stress-strain response predictions for uniaxial tension (using the material parameters from \citet{ma2026polymer}) over four decades of strain rate with the baseline Matlab results. We use a mass scaling factor (MSF) for the explicit simulations  (0.1 s$^{-1}$: MSF = $10^{6}$, 0.01 s$^{-1}$: MSF = $10^{8}$, 0.001 s$^{-1}$: MSF = $10^{10}$, 0.0001 s$^{-1}$: MSF = $10^{12}$) to increase the time step size and reduce the computation time. The predictions are stable and accurate even for very large MSFs.

\begin{figure*}[ht!]
    \begin{center}
		\includegraphics[width=0.95\textwidth]{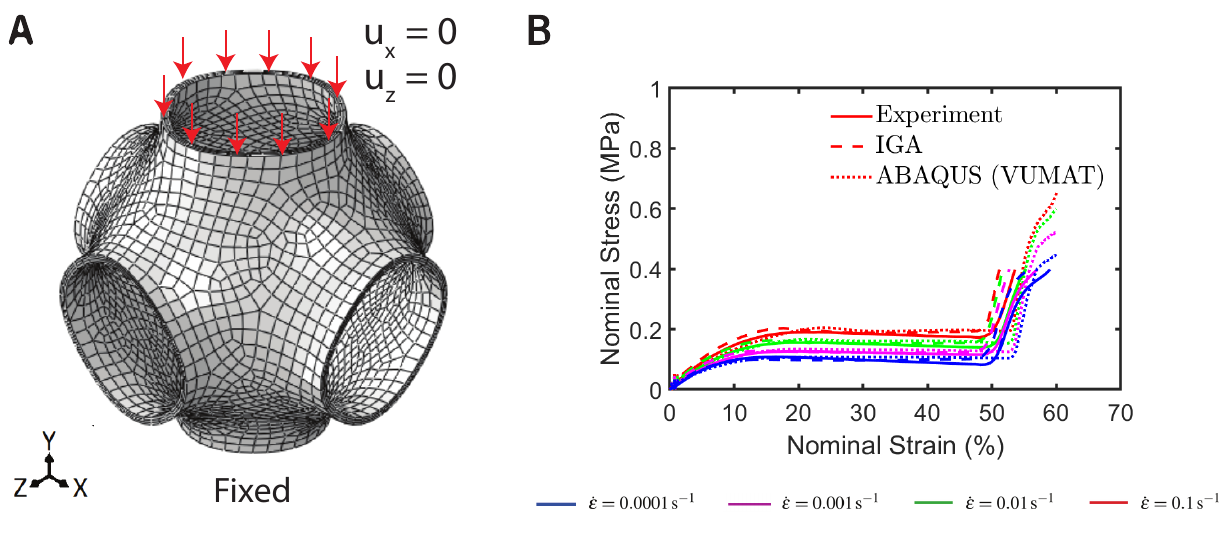}
	\end{center}
    \caption{\small {\textbf{Unit cell compression: Comparison of FEM predictions with experimental and baseline IGA results.} \textbf{(A)} FE mesh used for the unit cell with 6201 C3D8R elements and 3 elements through the thickness, along with the boundary conditions. \textbf{(B)} Comparison of FEM nominal stress-strain response predictions with experimental and baseline IGA (uniform thickness) results from \citet{ma2026polymer} over four decades of strain rates. %The following MSFs were used - 0.1 s$^{-1}$: MSF = $10^{6}$, 0.01 s$^{-1}$: MSF = $10^{8}$, 0.0001 s$^{-1}$: MSF = $10^{10}$, 0.00001 s$^{-1}$: MSF = $10^{12}$. Deviations from IGA in the densification stage can be attributed to the differences in the contact algorithms used.
    }} 
    \label{ucdeffig3}
\end{figure*}

Next, we modeled the compression of a unit cell (experimental configuration shown in \Cref{TPMSgeomfig1}(D)). Three elements were meshed through the thickness, providing a good trade-off between accuracy and computational cost. Self-contact (assumed frictionless) of the unit cell upon compression was modeled using the built-in Abaqus tools. \Cref{ucdeffig3}(A) shows the FE mesh with C3D8R elements, boundary conditions, and \Cref{ucdeffig3}(B) shows the comparison of the FEM predictions for the nominal stress-strain response with the experimental and baseline IGA (uniform thickness) results for the four strain rates (0.1 s$^{-1}$: MSF = $10^{6}$, 0.01 s$^{-1}$: MSF = $10^{8}$, 0.001 s$^{-1}$: MSF = $10^{10}$, 0.0001 s$^{-1}$: MSF = $10^{12}$). Excellent agreement is observed, with the mismatch between FEM and IGA results occurring only in the densification regime attributable to the differences in the contact algorithms used. 

%As a sanity check, these large MSFs were also applied for the single-element uniaxial tension test simulations discussed earlier. Good agreement with the baseline Matlab predictions confirms that the large MSFs do not introduce any numerical artifacts.

\begin{figure*}[ht!]
    \begin{center}
		\includegraphics[width=0.8\textwidth]{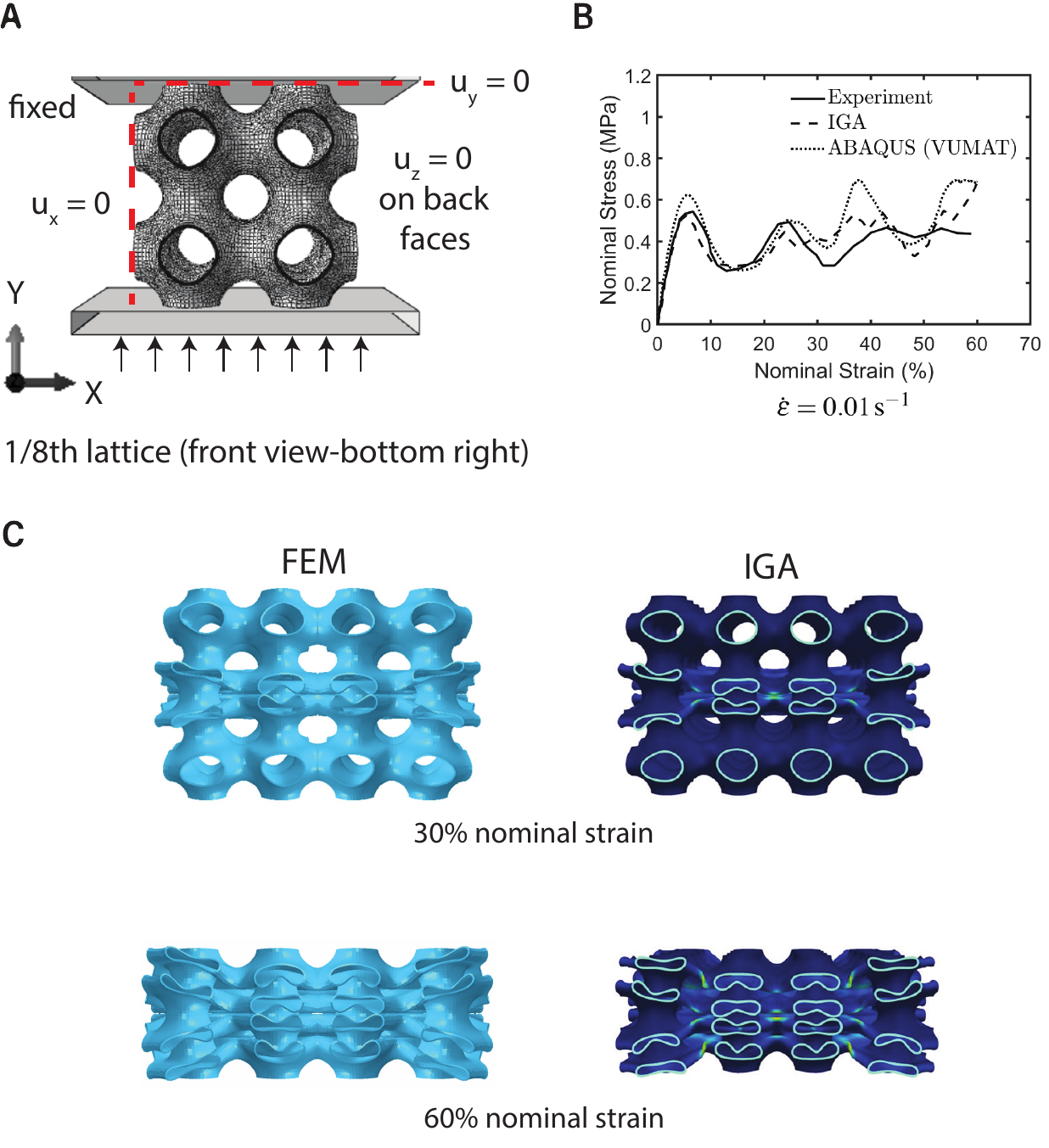}
	\end{center}
    \caption{\small {\textbf{$4\times4\times4$ lattice compression: Comparison of FEM predictions with experimental and baseline IGA results.} \textbf{(A)} FE mesh used for the 1/8 lattice (front view-bottom right) using the problem's symmetries with 44649 C3D8R elements and 3 elements through the thickness, along with the boundary conditions. The additional top rigid surface accounts for the contact of the 1/8 lattice with its symmetric counterpart above during deformation. \textbf{(B)} Satisfactory agreement of FEM (MSF = $10^{8}$) nominal stress-strain response predictions with experimental and baseline IGA (uniform thickness) results from \citet{ma2026polymer} for 0.01 s$^{-1}$ strain rate. Slight mismatch between FEM and IGA results after the initial softening regime can be attributed to differences in the contact algorithms used. \textbf{(C)} Strong qualitative agreement of the deformed lattice shapes (at 30~\% and 60~\% nominal strain) predicted by FEM with IGA results.}} 
    \label{ltdeffig4}
\end{figure*}

Further, we modeled the compression of a $4\times4\times4$ lattice [\Cref{TPMSgeomfig1}(E)]. The high mesh density in the unit cell shown in \Cref{ucdeffig3}(A) renders the full lattice simulation computationally expensive. Accordingly, only 1/8th of the lattice (front view-bottom right) was hence modeled using appropriate symmetry boundary conditions [see \Cref{ltdeffig4}(A)] to reduce the computation time. An appropriate modification to obtain physical results is the inclusion of a rigid surface at the top of the 1/8th lattice [see \Cref{ltdeffig4}(A)]. This accounts for the contact of the 1/8th lattice with its symmetric counterpart above during deformation. The bottom end block and the additional rigid surface were modeled as analytical rigid surfaces. All contacts were assumed to be frictionless and were modeled using the built-in Abaqus features. 
\Cref{ltdeffig4}(B) and (C) show the nominal stress-strain response at a representative strain rate of 0.01 $s^{-1}$, demonstrating good agreement between the FEM (C3D8R elements, MSF = $10^{8}$), the experimental results, and the baseline IGA (uniform thickness) simulations. Similar agreement was obtained for the other strain rates. The slight mismatch in the FEM and IGA nominal stress-strain curves after the initial softening regime can be attributed to differences in the contact algorithms used between the two computational platforms.

\subsection{Damage initiation criterion development and calibration}

\begin{figure*}[ht!]
    \begin{center}
		\includegraphics[width=0.80\textwidth]{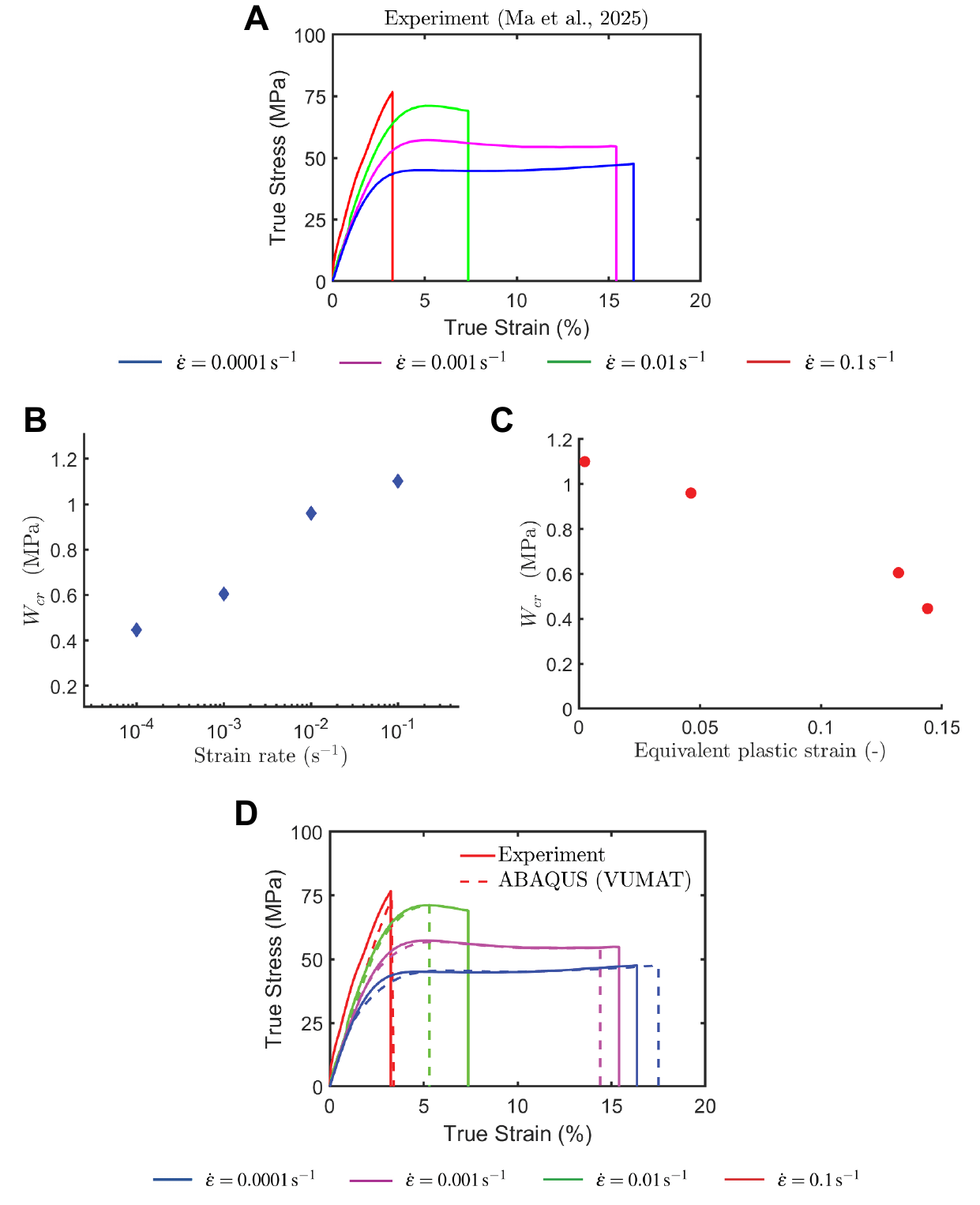}
	\end{center}
    \caption{\small {\textbf{Variation of $W_{cr}$ with key parameters and predictions for fracture under uniaxial tension.} \textbf{(A)} Experimental rate-dependent true stress-strain response of dogbone specimens under uniaxial tension till complete failure from \citet{ma2026polymer}. Variation of deformation model predicted $W_{cr}$ [evaluated using \Cref{Wedef} at experimental failure strains for each strain rate from \Cref{SWecrfig5}(A)] with \textbf{(B)} strain rate and \textbf{(C)} $\bar{\varepsilon}^p$. \textbf{(D)} Comparison of predictions [single C3D8R element computations] for failure under uniaxial tension using the deformation model and the fitted damage initiation criterion [\Cref{dgbnSWctcalbeq}, \Cref{dmgintcr}] with experimental results over the four decades of strain rate. Fracture was observed to be instantaneous in the experiments. Hence, it was approximated to occur in the model predictions when the damage initiation criterion is satisfied.}} 
    \label{SWecrfig5}
\end{figure*}

Since the deformation constitutive model for amorphous polymers in \citet{ma2026polymer} is developed in the rate/incremental form, a free energy density function ($\psi$) cannot be directly defined, and hence the $\psi$ based damage initiation criterion for polymers \citep{konale2025modeling} cannot be used directly. However, stored elastic energy can be obtained using the elastic part of stress work ($W^e$) defined as:

%\clearpage

%\setcounter{equation}{0}

\begin{equation}
\begin{split}
    &W^{e+} = \int^{t}_{0} J\textbf{T}^{+}_{\text{o}}:\textbf{D}^{e} dt,\\
    &\textbf{T}^{+}_{\text{o}} = \begin{cases}
\begin{aligned}
       \textbf{T}_{\text{o}} \hspace{0.88in} &&\text{if} \:\: J>1,\\
       \textbf{T}_{\text{o}} - \dfrac{\text{tr}(\textbf{T}_{\text{o}})}{3} \textbf{1} \hspace{0.1in} &&\text{if} \:\: J\le 1,
\end{aligned}
  \end{cases}
\end{split}
\label{Wedef}
\end{equation}

\noindent where $J, \textbf{T}_{\text{o}}$, and $\textbf{D}^{e}$ are the determinant of the deformation gradient tensor \textbf{F}, undamaged Cauchy stress tensor, and elastic part of the stretching tensor \textbf{D} = $\text{sym}(\dot{\textbf{F}} \textbf{F}^{-1})$, respectively. The superscript `$+$' indicates the distortional and tensile dilatational part of the corresponding quantity. This split of $W^{e}$ and $\textbf{T}_{\text{o}}$ ensures that damage is not driven by negative dilatational deformation and does not degrade the corresponding stiffness at a material point [\,$\textbf{T} = g(d) \textbf{T}^{+}_{\text{o}} + \textbf{T}^{-}_{\text{o}}, \:\: \text{`$-$' indicates negative dilatational part}, \:\: \text{degradation function} \:\:\: g(d) = (1-d)^2$\,]. Damage initiation can then be predicted using $W^{e+}$. The experimental rate-dependent true stress-strain response of dogbone specimens under uniaxial tension till complete failure is shown in \Cref{SWecrfig5}(A). The deformation model predicted critical values of $W^{e+}$: $W_{cr}$ [evaluated using \Cref{Wedef} at experimental failure strains for each strain rate from \Cref{SWecrfig5}(A)] vs. strain rate curve is shown in \Cref{SWecrfig5}(B). $W_{cr}$ is not constant and increases with strain rate. The variation of $W_{cr}$ with another variable of interest, i.e., the equivalent plastic strain $\bar{\varepsilon}^p$, is shown in \Cref{SWecrfig5}(C). For amorphous polymers, plastic strain can be one of the physical quantities affecting damage initiation \citep{narayan2021fracture}. It is observed that $W_{cr}$ decreases with increasing $\bar{\varepsilon}^p$. A possible physical interpretation is that as plastic strain accumulates, the polymer microstructure becomes progressively degraded through irreversible mechanisms such as molecular rearrangement, chain disentanglement, shear-banding, crazing, and microvoid formation or growth in additively manufactured polymer. This plastic strain based weakening lowers the energetic barrier for damage initiation, so the critical elastic energy required to break or separate local load-bearing molecular structures decreases.

%A possible physical interpretation is that as the plastic strain increases, it creates local weakening in the material which reduces the critical elastic energy needed to break the local load-bearing bonds for damage initiation. 

We propose a simple linear relation between $W_{cr}$ and $\bar{\varepsilon}^p$ as:
\begin{equation}
W_{cr} = A\, - B\, \bar{\varepsilon^p} \,,
\label{dgbnSWctcalbeq}
\end{equation}
where, $A$ and $B$ are damage initiation constants with $A$ representing the maximum possible energy per unit volume needed to break the local molecular structure and initiate local damage (maximum possible resistance), and $B$ determines the rate of decrease of the energy with increasing plastic strain. The model indicates that the additively manufactured polymer will fail before the equivalent plastic strain reaches the value $\frac{A}{B}$. From the fit of the experimental data, we obtain
%
% \begin{equation}
% \begin{split}
% \begin{cases}
% \begin{aligned}
%      A = 1.13 \;  \text{MPa}, \quad B = 4.39 \; \text{MPa}  &&\text{for} \:\: J>1
%    % A = 0.94 \;  \text{MPa}, \quad B = 3.65 \; \text{MPa}  &&\text{for} \:\: J\le 1.
% \end{aligned}    
% \end{cases}
% \end{split}
% \label{dgbnSWctcalbeq2}
% \end{equation}
%
\begin{equation}
     A = 1.13 \;  \text{MPa}, \quad B = 4.39 \; \text{MPa}.  
   % A = 0.94 \;  \text{MPa}, \quad B = 3.65 \; \text{MPa}  &&\text{for} \:\: J\le 1.
\label{dgbnSWctcalbeq2}
\end{equation}

% \begin{equation}
% \begin{split}
% &W^{e+}_{cr} =\begin{cases}
% \begin{aligned}
%      -4.39 \, \bar{\varepsilon^p} + 1.13 \:\: \text{MPa} \hspace{0.05in} &&\text{if} \:\: J>1,\\
%     -3.65 \, \bar{\varepsilon^p} + 0.94 \:\: \text{MPa} \hspace{0.05in} &&\text{if} \:\: J\le 1.
% \end{aligned}    
% \end{cases}
% \end{split}
% \label{dgbnSWctcalbeq}
% \end{equation}

The damage initiation criterion can then be written as 
\begin{equation}
\text{Damage initiates when:} \:\:
W^{e+} = W_{cr}\,.
\label{dmgintcr}
\end{equation}

% For loading conditions with $J>1$, both the total elastic energy and the distortional (deviatoric) energy are evaluated, and their critical values are determined at fracture. In contrast, for $J\leq1$, the compressive dilatational contribution is not considered damage-driving, since compressive states tend to close defects rather than promote crack growth. Therefore, only the distortional component is assumed to contribute to damage. Assuming that the critical distortional work is transferable across loading modes \citep{chaouadi1994damage}, the $J\leq1$ failure criterion can be defined using the critical distortional work obtained from the $J>1$ tensile tests. This leads to slightly smaller values of the parameters $A$ and $B$, since the critical work $W_{cr}$ now consists solely of distortional energy.
% %
% \begin{equation}
% \begin{split}
% \begin{cases}
% \begin{aligned}
%      A = 0.94 \;  \text{MPa}, \quad B = 3.65 \; \text{MPa}  &&\text{for} \:\: J\leq 1
% \end{aligned}    
% \end{cases}
% \end{split}
% \label{dgbnSWctcalbeq22}
% \end{equation}

\noindent Damage initiation criterion based on a combination of $W^{e+}$ and $\bar{\varepsilon}^{p}$, like the one proposed here, could apply to other amorphous glassy polymers as well. \Cref{SWecrfig5}(D) shows the comparison of predictions (single C3D8R element computations) for failure under uniaxial tension using the deformation model and the fitted damage initiation criterion [\Cref{dgbnSWctcalbeq}, \Cref{dmgintcr}] with experiments. The experiments show that the failure, when it  occurs, is instantaneous. Therefore the occurrence of fracture in the experiments is used for the damage initiation criterion. 

\subsection{Damage growth modeling}

The standard gradient-damage framework \citep{konale2025modeling, anand2026gradient} specialization can be directly used by replacing $\psi^{+}$ with $W^{e+}$ as shown below
\begin{equation}
\begin{split}
    \\& \hspace{0 cm} \mathcal{H}(t) \overset{\text{def}}{=} \underset{s \in [0,t]}{\text{max}} \Big[\langle W^{e+}(s) - W_{cr} \rangle\Big],   
 \\& \langle W^{e+}(s) - W_{cr} \rangle = \begin{cases}
\begin{aligned} 
			&0 \hspace{0.2in} \hspace{0.1in} &&\text{if}  \hspace{0.1in} W^{e+}(s) - W_{cr} < 0,\\
			& W^{e+}(s) - W_{cr}  && \text{if} \hspace{0.1in} W^{e+}(s) - W_{cr} \geq 0, 
\end{aligned}
	\end{cases} 
\\& \zeta \dot{d} = 2(1-d)\mathcal{H} -2\Gamma(d- l^{2} \Delta d),
\end{split}
\label{monotonically increasing history function app}
\end{equation}

\noindent where $d$ is the damage field, $d \in [0,1]$ with $d=0$ represents undamaged and $d=1$ represents completely damaged material point. $\mathcal{H}$ is the referential monotonically increasing damage driving history function, $l$ is a length scale parameter for the diffuse damage zone, $\zeta$ is a constant kinetic modulus determining the timescale of damage growth, and $\Gamma$ is a coefficient with units of energy per unit volume representing a part of the dissipated energy density after damage initiates and until it reaches $d=1$. The deformation model from \citet{ma2026polymer}, the proposed damage initiation criterion [\Cref{dgbnSWctcalbeq}, \Cref{dmgintcr}], and damage growth modeling [\Cref{monotonically increasing history function app}] constitute the damage and fracture constitutive model for polymeric TPMS lattices presented in this work. 

The values of the damage growth related parameters are taken as $l = 2$~mm, $\zeta = 1$~kPa$\cdot$s, $\Gamma = 4$~kPa. The parameter $l$ is treated as a numerical regularization parameter and is chosen as equal to 5 times the typical element size to avoid mesh dependency. The VUMAT (user material subroutine) - VUEL (user element subroutine) - VUSDFLD (subroutine to pass information between VUMAT and VUEL) subroutines approach for Abaqus/Explicit to solve for displacement [using VUMAT] and damage [using VUEL] degrees of freedom, discussed in \citet{konale2025modeling} is used here. The parameters $\zeta$, $\Gamma$ were fit to the parts of stress-strain curves corresponding to fracture in \Cref{SWecrfig5}(A). It can be noted that due to the experimentally observed almost instantaneous damage growth (very rapid drop in experimental load curves at failure), unique values for $\zeta$ and $\Gamma$ cannot be obtained. Any reasonably low values of $\zeta$ leading to rapid damage growth and low values of $\Gamma$ compared to $W_{cr}$ [energy density dissipated till damage initiation ($W_{cr}$) is much larger than the energy that dissipates after damage initiation ($\Gamma$) to complete failure] can represent the experimental observations.

\subsection{Damage initiation calibration: unit cell}

While 3D-printed dog-bone specimens and unit cells share the same base polymer, their mechanical responses can vary due to geometry-specific subtle variations in the printing process. Recalibrating the damage initiation criterion parameters using unit-cell experiments can account for such geometry-dependent effects and improve the model's overall predictive ability. This exercise has been performed for the deformation-related model parameters in \citet{ma2026polymer}, and the corresponding values have been used for unit cell and lattice computations here. Once an element is completely damaged, its stiffness is degraded to nearly zero (a small conditioning number is usually added for numerical stability), and it can undergo severe deformation. This is particularly relevant in multi-element numerical simulations where a single severely distorted element can cause simulation termination. Also, due to the explicit time integration used, damage values can sometimes be nonphysical. Hence, $d$ from VUEL is processed as follows before using it in VUMAT to degrade the undamaged stress:

\begin{equation}
d^{\textrm{VUMAT}} = \begin{cases}
\begin{aligned}
       0.95 \hspace{0.2in} &\text{if} \hspace{0.2in} d^{\textrm{VUEL}}>0.95,\\
       0 \hspace{0.2in} &\text{if} \hspace{0.2in} d^{\textrm{VUEL}}<0,
\end{aligned}
  \end{cases} 
\end{equation}

\noindent with the subscripts $\textrm{VUMAT}$, $\textrm{VUEL}$ distinguishing $d$ in the two subroutines. Note that the maximum value of $d^{\textrm{VUMAT}}$ is limited to 0.95, so that the element retains a small but sufficient stiffness to avoid severe distortion and aid in numerical convergence. This $d^{\textrm{VUEL}}$ processing is applied alongside the default element distortion settings in Abaqus. The parameters in the assumed linear relationship between $W_{cr}$ and $\bar{\varepsilon}^p$ [\Cref{dgbnSWctcalbeq}] are re-calibrated (due to effects of grometry/shape on 3D-printed mtertial) to tensile experiments on unit cells and found to be
\begin{equation}
     A = 1.50 \;  \text{MPa}, \quad B = 1.78 \; \text{MPa}. 
\label{ucSWctcalbeq}
\end{equation}
% %
% \begin{equation}
% \begin{split}
% \begin{cases}
% \begin{aligned}
%      A = 1.50 \;  \text{MPa}, \quad B = 1.78 \; \text{MPa}  &&\text{for} \:\: J>1,\\
%     A = 0.86 \;  \text{MPa}, \quad B = 0.88 \; \text{MPa}  &&\text{for} \:\: J\le 1.
% \end{aligned}    
% \end{cases}
% \end{split}
% \label{ucSWctcalbeq}
% \end{equation}
%
% \begin{equation}
% \begin{split}
% &W^{e+}_{cr} =\begin{cases}
% \begin{aligned}
%      -1.78 \, \bar{\varepsilon^p} + 1.5 \:\: \text{MPa} \hspace{0.05in} &&\text{if} \:\: J>1,\\ -0.88 \, \bar{\varepsilon^p} + 0.86 \:\: \text{MPa} \hspace{0.05in} &&\text{if} \:\: J \le 1.
% \end{aligned}    
% \end{cases}    
% \end{split}
% \label{ucSWctcalbeq}
% \end{equation}

\begin{figure*}[ht!]
    \begin{center}
		\includegraphics[width=0.85\textwidth]{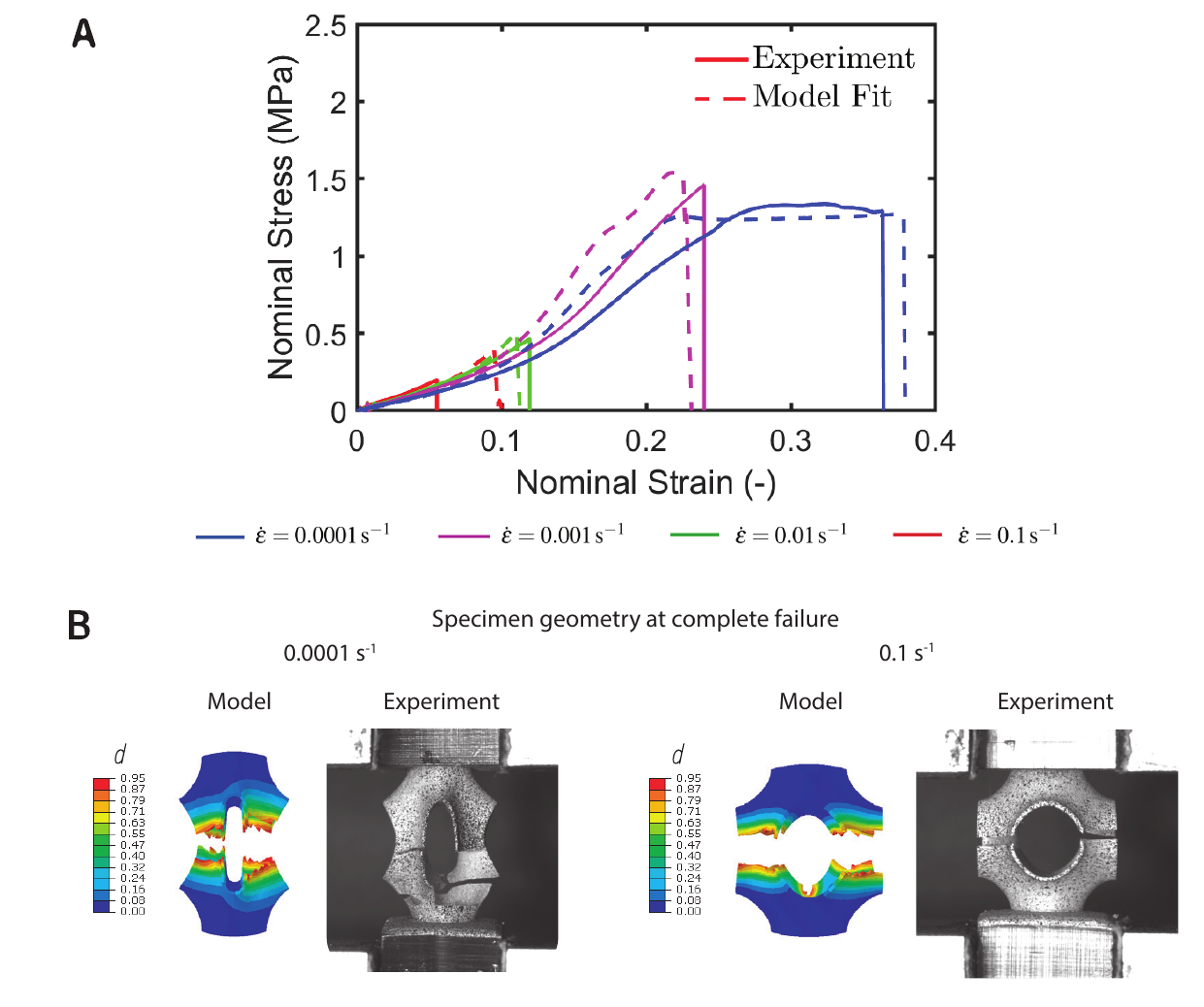}
	\end{center}
    \caption{\small {\textbf{Damage initiation criterion recalibration for unit cell and fracture model fit (unit cell under uniaxial tension).} %The 3D-printed dog-bone specimens and unit cells may exhibit slightly different material responses due to subtle variations in the printing process associated with different geometries. The damage initiation criterion parameters were hence recalibrated using unit-cell tension experiments to eliminate such geometry-dependent effects and improve the fracture model's overall predictive ability. 
    \textbf{(A)} The model's fit over the four decades of strain rate shown through comparison of nominal stress-strain response predictions till complete failure with experiments. The FE mesh with C3D8R elements and boundary conditions in \Cref{ucdeffig3}(A) were used. \textbf{(B)} Good qualitative agreement of model-predicted specimen geometry at complete failure and fracture locations with experimental results for 0.0001 s$^{-1}$ and 0.1 s$^{-1}$ strain rates. Elements with $d=0.95$ were removed for visualization.}} 
    \label{dmgucrecalbfig6}
\end{figure*}

\noindent The fracture model fit over the four decades of strain rate is shown in \Cref{dmgucrecalbfig6}(A) through the comparison of nominal stress-strain response predictions till complete failure with experimental results. The FE mesh with C3D8R elements and boundary conditions in \Cref{ucdeffig3}(A) were used. The fracture model predictions for specimen geometry at complete failure and fracture locations in \Cref{dmgucrecalbfig6}(B) show good qualitative agreement with the experimental results considering the variability and uncertainty of fracture measurements for polymers. Elements with $d=0.95$ were removed for visualization.

% Good qualitative agreement of model-predicted fracture locations with experimental results.

\subsection{Fracture model and numerical simulation capability validation}

%We now apply the calibrated fracture model and numerical simulation capability to unseen geometries and loading scenarios involving inhomogeneous three-dimensional deformations ($4\times4\times4$ lattices subjected to tension) and compare the results with experimental data for robust validation. Figure (A) shows the model-predicted nominal stress-strain curves (for a $4\times4\times4$ lattice subjected to tension) till complete failure, showing good agreement with the experimental results over the four decades of strain rate. The specimen shapes at complete failure, and the fracture locations also qualitatively agree with the experiments, as shown in Figure (B). Elements with $d=0.95$ were removed for easier visualization. The fracture model, hence, provides reasonably accurate quantitative predictions while capturing the key physical phenomena. 

We now apply the recalibrated fracture model and numerical simulation capability to unseen geometries and loading scenarios involving inhomogeneous three-dimensional deformations and compare the results with experimental data for validation. C3D8R elements were used for all simulations reported in this section. It can be noted that experimental fracture measurements for polymers exhibit significant variability and uncertainty. Variations due to the printing process can exacerbate this issue for larger structures (lattices). 

%We now apply the recalibrated fracture model and numerical simulation capability to unseen geometries and loading scenarios involving inhomogeneous three-dimensional deformations and compare the results with experimental data for robust validation. 

\subsubsection{Unit cell subjected to uniaxial non-monotonic loading-unloading}

\begin{figure*}[ht!]
   \begin{center}
		\includegraphics[width=0.7\textwidth]{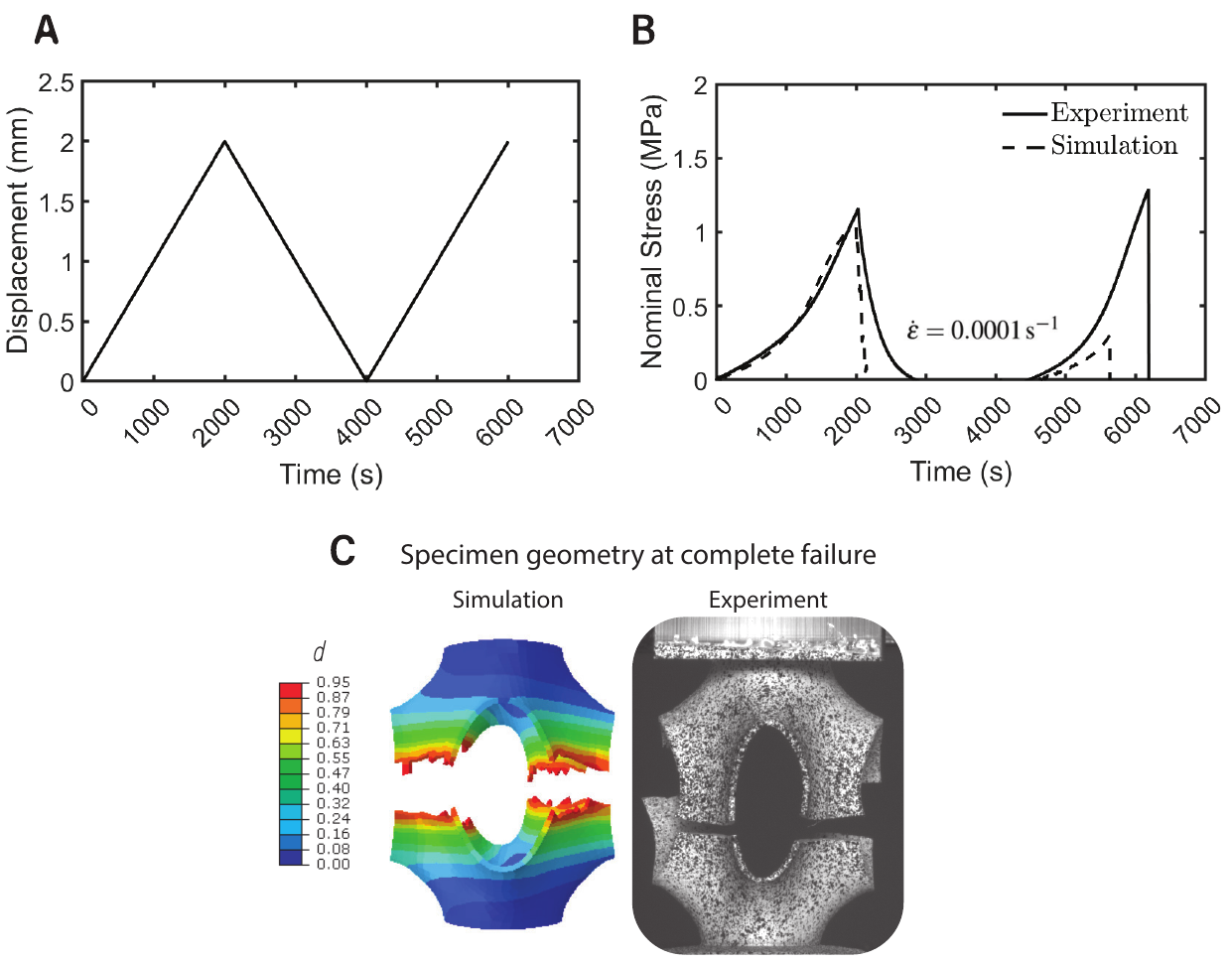}
	\end{center}
    \caption{\small {\textbf{Fracture model validation: unit cell under uniaxial non-monotonic loading.} \textbf{(A)} Non-monotonic loading profile. Comparison between \textbf{(B)} quantitative and \textbf{(C)}  qualitative predictions of the fracture model with the experimental results.% (considering the variations in experimental results for polymer fracture). The FE mesh with C3D8R elements and boundary conditions in \Cref{ucdeffig3}(A) were used. Elements with $d=0.95$ were removed, and positive stress values are plotted for visualization. The proposed damage initiation criterion captures damage dependence on both stored elastic energy and plastic strain.
    }} 
   \label{ucnonmonovalidationfig7}
\end{figure*}

We consider a unit cell subjected to a uniaxial non-monotonic loading-unloading-reloading profile shown in \Cref{ucnonmonovalidationfig7}(A). The FE mesh and boundary conditions in \Cref{ucdeffig3}(A) were used. Quantitative and qualitative comparisons (considering the variability in polymer fracture experimental results discussed earlier) between the fracture model and the experimental results are shown in \Cref{ucnonmonovalidationfig7}(B) and (C). Elements with $d=0.95$ were removed, and positive stress values are plotted for visualization. The damage initiation criterion, hence, captures the dependence of damage on both stored elastic energy and plastic strain.  

%The fracture model, hence, provides reasonably accurate quantitative predictions while capturing the key physical phenomena. 

\subsubsection{$2\times2\times2$ lattices subjected to uniaxial tension}
\label{2by2by2unaxsec}

\begin{figure*}[ht!]
    \begin{center}
		\includegraphics[width=0.8\textwidth]{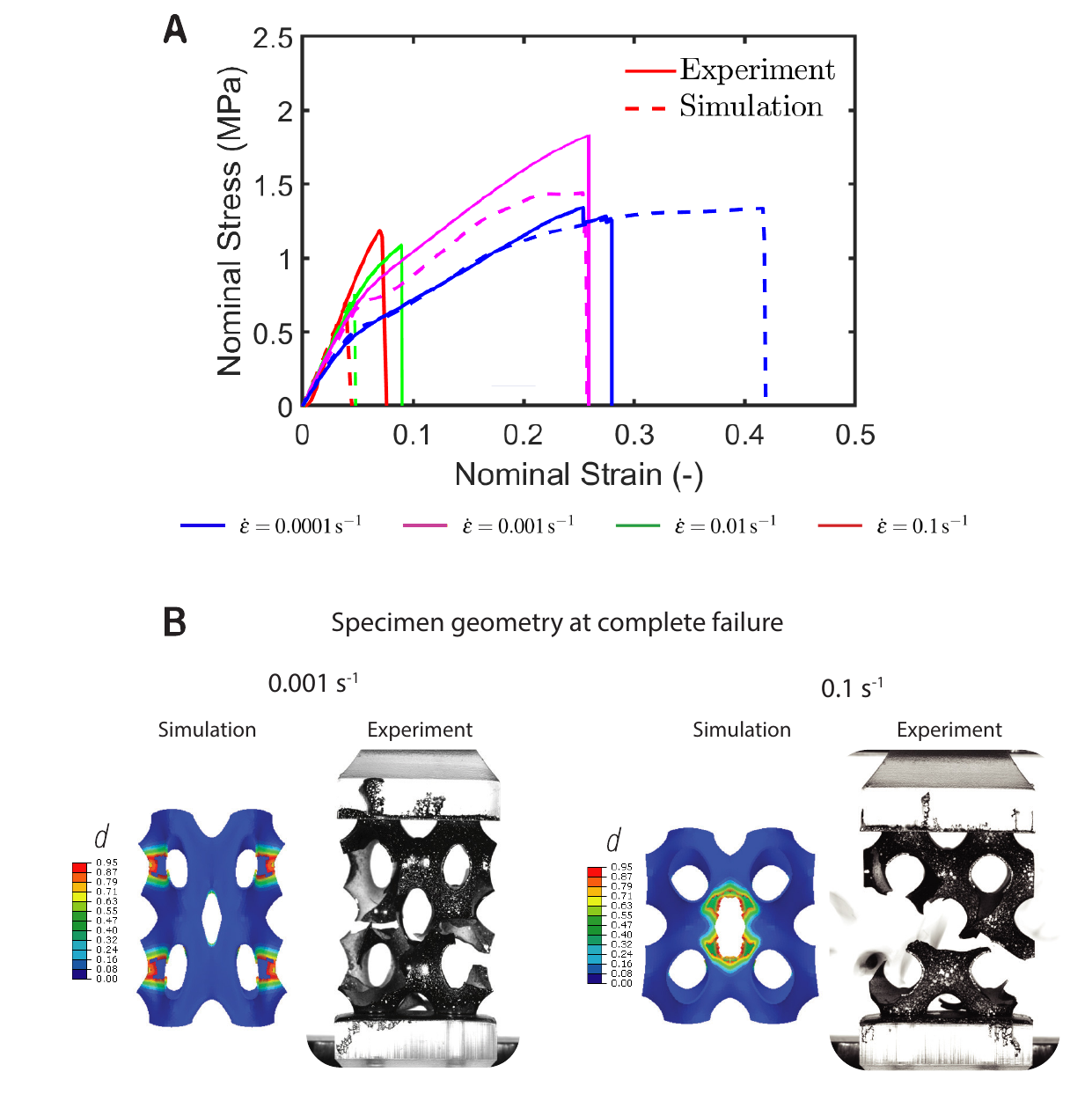}
	\end{center}
    \caption{\small {\textbf{Fracture model validation: $2\times2\times2$ lattice under uniaxial tension.} \textbf{(A)} Close agreement of the fracture model-predicted nominal stress-strain curves till complete failure with the experimental results. The deviation at the lowest strain rate is expected, as the larger deformation attained prior to failure increases the influence of printing defects, which can lead to premature failure. %The FE mesh in \Cref{ucdeffig3}(A) was used for a unit cell as the 1/8th part (front view-bottom right) of the lattice with appropriate symmetry boundary conditions [see \Cref{ltdeffig4}(A)]. 
    \textbf{(B)} The model correctly predicts specimen shapes at complete failure along with fracture locations for 0.001 s$^{-1}$ and 0.1 s$^{-1}$. %Elements with $d=0.95$ were removed for visualization. Along with providing reasonably accurate quantitative predictions, the fracture model captures the key physical phenomenon.
    }} 
   \label{ltsmallvalidationfig8}
\end{figure*}

A lattice system ($2\times2\times2$) was first considered due to its lower susceptibility to fabrication-induced variability than the larger $4\times4\times4$ lattice. The FE mesh in \Cref{ucdeffig3}(A) was used for a unit cell as the 1/8th part (front view-bottom right) of the lattice with appropriate symmetry boundary conditions [see \Cref{ltdeffig4}(A)]. \Cref{ltsmallvalidationfig8}(A) shows the model-predicted nominal stress-strain curves to complete failure. The predictions show reasonably good agreement with the experimental results. For the slowest strain rate, the higher level of deformation till failure shows only a reasonable agreement, which can be attributed to variability within fabricated samples and the variability playing a larger role when the samples span larger stretches. The specimen shapes at complete failure, and the fracture locations also qualitatively agree reasonably well with the experiments, as shown in \Cref{ltsmallvalidationfig8}(B) for 0.001 s$^{-1}$ and 0.1 s$^{-1}$. Elements with $d=0.95$ were removed for visualization. The fracture model, hence, provides reasonably accurate quantitative predictions while capturing the key physical phenomena.

\subsubsection{$4\times4\times4$ lattices subjected to uniaxial tension}
\label{4by4by4ltunxvald}

\begin{figure*}[ht!]
    \begin{center}
		\includegraphics[width=0.65\textwidth]{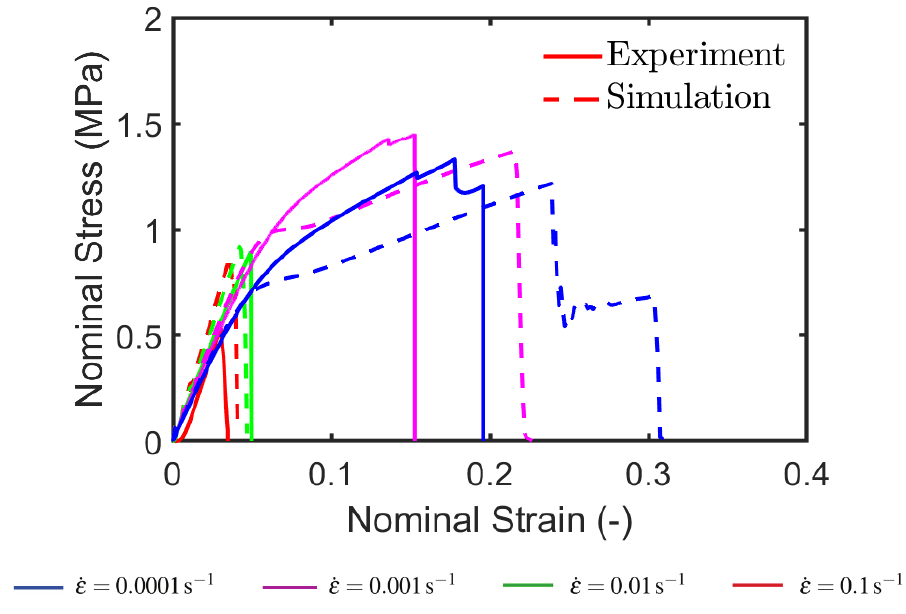}
	\end{center}
    \caption{\small {\textbf{Fracture model validation: $4\times4\times4$ lattice under uniaxial tension.} Reasonable agreement (considering the additional variations due to the printing process across the relatively large lattice) of the fracture model-predicted nominal stress-strain curves till complete failure with the experimental results over the four decades of strain rate. %The FE mesh and boundary conditions in \Cref{ltdeffig4}(A) were used. At the two slower strain rates, only a reasonable agreement can be expected following the discussion in \Cref{2by2by2unaxsec}.
    }} 
   \label{ltvalidationfig9}
\end{figure*}

\Cref{ltvalidationfig9} shows the model-predicted nominal stress-strain curves till complete failure for $4\times4\times4$ lattices subjected to uniaxial tension. The FE mesh and boundary conditions in \Cref{ltdeffig4}(A) were used. Reasonably good agreement is observed for the two higher strain rates, as the deformation to fracture is relatively small, and the probability of printing-related variations for this larger lattice affecting the experimental results is minimal. On the other hand, for the two slower strain rates, the simulations predict the right overall features and trends but the accuracy is reduced due to large deformation and large sized lattice resulting in more printing-related experimental variations. 

%The specimen shape at complete failure also agrees qualitatively with the experiment, as shown in \Cref{ltvalidationfig9}(B) for 0.1 s$^{-1}$. Elements with $d=0.95$ were removed for visualization.

\section{Conclusions}

Polymeric triply periodic minimal surface (TPMS) architected materials and structures offer an efficient route for obtaining superior mechanical performance while maintaining low density and environmental stability. The scientifically and practically important damage and fracture modeling of additively manufactured polymers is still lacking in the literature. We attempt to fill this critical gap by developing and numerically implementing damage and fracture modeling approach for additively manufactured polymeric TPMS.  

We began by implementing the rate-dependent viscoplastic deformation constitutive model for amorphous polymers from \citet{ma2026polymer} in the finite element software Abaqus/Explicit through a user material subroutine. The numerical implementation is robustly tested through single-element (uniaxial tension of dogbone specimens) and multi-element (unit cell and lattice under uniaxial compression) simulations, and by comparing the predictions with the baseline Matlab, isogeometric analysis (IGA), and experimental results. Next, we proposed a damage initiation criterion for amorphous polymers based on a measure of stored elastic energy density (elastic part of stress work) and equivalent plastic strain. We then implemented the model in a gradient-damage framework in Abaqus and applied the combined deformation-damage model and simulation capability to three-dimensional, inhomogeneous deformation problems, including a unit cell subjected to uniaxial non-monotonic loading and different lattice structure sizes under tension. The model was validated through quantitative and qualitative agreement with experimental results. The damage model and numerical simulation capabilities such as the one presented here can enable \textit{in silico} designs of polymer-based architected materials and structures.

\section*{Acknowledgments}
The authors gratefully acknowledge support from the Office of Naval Research (ONR), USA, under Grant Nos. N00014-26-1-2135, N00014-21-1-2815, and N00014-23-1-2688, as well as partial support from the Hibbitt Postdoctoral Fellowship awarded to A. Gupta and K. Alkhoury at the Brown University School of Engineering.

\section*{Declaration of Interest}
The authors declare no competing interests.

\begin{appendices}

\clearpage

\section*{Appendix}

\subsection*{STL file conversion to an independently meshable solid body in Abaqus}

As the specimens were fabricated using SLA, STL files are the starting point for geometry. Continuum elements were used, keeping in mind the restrictions of the Abaqus/Explicit FE software-based damage modeling approach in \citet{konale2025modeling} used here. Hence, the STL files have to be converted to an independently meshable solid body. We lay out the conversion process here using Autodesk Fusion 360 for the unit cells and lattices:\\ 

\textbf{\underline{Unit cell}}
\begin{itemize}
    \item Import the STL file in Fusion 360
    \item Go to `Mesh' module - `Prepare' - `Generate Face groups'
    \item `Modify' - `Convert Mesh' - Method: `Organic' - Choose conversion accuracy as needed
    \item Export solid body as STEP file
    \item Import STEP file into Abaqus
    \item Go to `Mesh' module in Abaqus - `Virtual Topology' - `Combine faces' - select the entire part, follow the prompts
    \item Now the part can be meshed using the standard procedure in Abaqus
\end{itemize}
 
\textbf{\underline{Lattice}}
\begin{itemize}
    \item Import the STL file in Fusion 360
    \item Reduce number of triangular elements if needed to reduce conversion time, `Mesh' module - `Reduce'
    \item `Repair' - `Stitch and Remove'
    \item `Modify' - `Convert Mesh' - Method: `Organic' - Choose conversion accuracy as needed
    \item `Solid' - `Split Body' - Create datum planes to split each unit cell into eight equal parts. Partitioning in Abaqus is impractical due to the very large computation time 
    \item Export solid body as STEP file 
    \item Import STEP file into Abaqus, Choose `Combine into single part', `Merge solid regions', `Retain intersection boundaries' in the dialog box
    \item Go to `Mesh' module in Abaqus - `Virtual Topology' - `Combine faces' - select the entire part, follow the prompts
    \item There will be a 1/8th part of a unit cell that is not assigned sweep mesh control. Locate it, go to `Tools' - `Partition' - `Face' - Select the red face - Use `Shortest distance between two points' - Select the two points on that face which are not connected by a line. Now the sweep mesh control will be assigned to it
    \item Now the part can be meshed using the standard procedure in Abaqus

%If prompted to repair the geometry, follow the prompts

\end{itemize}

\end{appendices}

\clearpage

\bibliographystyle{abbrvnat}
\bibliography{BibtexFile}

\end{document}